Title: Correlated Insulator and Chern Insulators in Pentalayer Rhombohedral Stacked Graphene

Author list: Tonghang Han[1]†, Zhengguang Lu[1]†, Giovanni Scuri[2,3], Jiho Sung[2,3], Jue Wang[2,3], Tianyi Han[1], Kenji Watanabe[4], Takashi Taniguchi[5], Hongkun Park[2,3], Long Ju[1]*

Affiliations: [1]Department of Physics, Massachusetts Institute of Technology, Cambridge, MA, USA.

[2]Department of Chemistry and Chemical Biology, Harvard University, Cambridge, MA, USA.

[3]Department of Physics, Harvard University, Cambridge, MA, USA.

[4]Research Center for Electronic and Optical Materials, National Institute for Materials Science, 1-1 Namiki, Tsukuba 305-0044, Japan

[5]Research Center for Materials Nanoarchitectonics, National Institute for Materials Science, 1-1 Namiki, Tsukuba 305-0044, Japan

*Corresponding author. Email: longju@mit.edu †These authors contributed equally to this work.

**Abstract:**

**Rhombohedral stacked multilayer graphene hosts a pair of flat bands touching at zero energy, which should give rise to correlated electron phenomena that can be further tuned by an electric field. Furthermore, when electron correlation breaks the isospin symmetry, the valley-dependent Berry phase at zero energy may give rise to topologically non-trivial states. Here, we measure electron transport through hBN-encapsulated pentalayer graphene down to 100 mK. We observed a correlated insulating state with resistance R>MΩ at charge density $n$=0 and displacement field $D$=0. Tight-binding calculations predict a metallic ground state under these conditions. By increasing $D$, we observed a Chern insulator state with $C$ = -5 and two other states with $C$ = -3 at magnetic field around 1 T. At high $D$ and $n$, we observed isospin-polarized quarter- and half-metals. Hence, rhombohedral pentalayer graphene exhibits two different types of Fermi-surface instabilities, one driven by a pair of flat bands touching at zero energy, and one induced by the Stoner mechanism in a single flat band. Our results establish rhombohedral multilayer graphene as suitable system to explore intertwined electron correlation and topology phenomena in natural graphitic materials without the need for moiré superlattice engineering.**

**Main text:**

Van der Waals heterostructures of two-dimensional materials have been actively studied to engineer novel device structures and physical properties. In recent years, the introduction of moiré superlattices between adjacent two-dimensional materials has led to a rich spectrum of correlated and topological electron phenomena, including Mott insulators, superconductivity and (quantum) anomalous Hall effects[1–16]. These studies often rely on specific combinations of materials and twist angles — some complex engineering to realize the wanted phenomena. It is intriguing to ask if one can get electron correlation and topology in crystalline two-dimensional materials without moiré effects. The answer to this question not only clarifies the exact role of moiré superlattice in already discovered phenomena, but could also provide a more general recipe to get similar phenomena in other condensed matter systems where it is hard to introduce moiré.

Multilayer graphene in the rhombohedral stacking order is potentially such a platform, where a large density-of-states (DOS) at zero energy derives from a pair of flat conduction and valence bands. Simultaneously, these flat bands carry large valley-dependent Berry phase due to the pseudo-spin winding around the Dirac point[17–21]. As a result, both correlated and topological states could emerge from this material without any moiré superlattice. However, previous experiments on hBN-encapsulated devices showed low resistivity at ~kΩ down to milli-Kelvin temperatures at zero displacement field $D$ and zero charge density $n$, which do not necessarily invoke a correlated electron picture[22–24]. Scanning tunneling spectroscopy studies on rhombohedral-stacked regions of twisted double-bilayer graphene did not show a clean gap either[25,26]. Experiments using suspended graphene devices indeed observed an interacting-induced insulating state, but the measurements and data interpretation suffered from either the two-terminal geometry or the very limited parameter space. Crucially, topological states with clear signatures of quantized longitudinal resistance $R_{xx}$ and Hall resistance $R_{xy}$ values have not been identified[27–31].

Here we overcame these obstacles by employing an hBN-encapsulated dual-gated pentalayer graphene in the rhombohedral stacking order (Fig. 1a). We purposely avoid the moiré effect by choosing a large twist angle between the graphene and neighboring hBN layers. As shown in Fig. 1b, the Hall-bar device configuration allows for accurate measurements of $R_{xx}$ and $R_{xy}$, while its non-suspended geometry enables access to a wide range of charge density $n$ and displacement field $D$. At zero $D$ and correspondingly zero interlayer potential difference $\Delta$, our tight-binding calculation predicts flat conduction and valence bands near zero energy (Fig. 1c), from which a large DOS emerges (Fig. 1d). This fact points to a possibility of Fermi-surface instability different from that driven by the Stoner mechanism in a single band. The latter can also be observed in our device at large $D$, as we will show in the following. Compared to layer number

$N \leq 3$, the DOS in pentalayer graphene is larger due to the $E \propto \pm |k|^N$ energy dispersion of rhombohedral-stacked graphene (Fig. 1d), where $k$ is the electron momentum. In even thicker layers, the screening effect increases with increased $N$ and counteracts the effect of flat bands[31–33]. As a result, $N = 5$ is likely to be close to an ideal thickness for observing electron correlation at zero energy.

**Phase diagram of rhombohedral-stacked pentalayer graphene**

Figure 1e & f show the 2D color plot of the device resistance $R_{xx}$ by tuning $n$ and $D$ at an out-of-plane magnetic field $B = 0$ T and 2 T, respectively. The measurements are taken at a temperature of $T = 100$ mK. Combining both plots, we made the following three observations:

Firstly, an insulating state with $R_{xx} = 10$ MΩ appears at $D = n = 0$. This resistance is three orders-of-magnitude higher than that in both thinner and thicker hBN-encapsulated rhombohedral-stacked graphene devices[22–24]. This insulator state is clearly beyond the prediction of tight-binding calculation and is a manifestation of gap-opening due to electron correlation at zero energy. We will call it a correlated insulator (CI) throughout the manuscript. The insulating state at $D > 0.19$ V/nm is likely a band insulator (BI), as the interlayer potential $\Delta$ will dominate the Hamiltonian at large $D$.

Secondly, three states with $R_{xx}$ close to zero emerge in a magnetic field, as indicated by dashed circles in Fig. 1f. As we will show more data and discuss in detail later, these states show quantized $R_{xy}$ and we conclude that they are Chern insulators (CHI). They are located at finite doping and $D = 0.1 - 0.19$ V/nm, neighboring states with $R_{xx} \sim$ kΩ at $n = 0$. The CHI states have a strong connection with the CI state, and they will be the focus of our following discussions.

Thirdly, isospin-polarized metallic states exist at finite $D$ and $n > 0.5*10^{12}$ cm$^{-2}$. At $B = 0$ T, three regions of the $n$-$D$ plot can be identified as un-polarized full-metal (UP, the magenta dot), spin-polarized half-metal (SPHM, the yellow dot) and isospin-polarized quarter-metal (IPQM, the red dot). The degeneracy of these states can be seen from the period of quantum oscillations in Fig. 1f (or more rigorously analyzed following the so-called Fermiology[23,24]). The spin and valley characters of these states can be deduced from the anomalous Hall resistance (see Extended Data Figure 4). We note that an elliptical-shaped region in Fig. 1f (the second red dot) also corresponds to IPQM which is only revealed under a finite magnetic field.

These observations demonstrate a rich collection of electron correlation and topological effects in pentalayer graphene, driven by two distinct mechanisms. The first mechanism originates from the co-existence of flat electron and hole bands near zero energy, resulting in a correlated insulating state. Previous literature suggested a layer anti-ferromagnetic (LAF) state as the ground state[19,30]. The second one is the Stoner-type instability incurred from a single band with large DOS, which can be created by opening up the gap of rhombohedral multilayer graphene at large $D$. The resulted SPHM and IPQM states have also been observed in bilayer and trilayer graphene[23,34–36]. Pentalayer graphene, however, shows both types of correlation effects. We will focus on phenomena due to the first mechanism in the following.

**Correlated insulator at zero $D$ and zero $n$**

In Fig. 1d, the single-particle calculation reveals a significantly greater DOS at $D = n = 0$ in pentalayer graphene than those in bilayer and trilayer graphene. In the latter systems, low resistances were observed at $D = n = 0$, which aligns with the semi-metallic states predicted by the tight-binding calculations[23,24,34–36]. The resistance of pentalayer graphene at $D = n = 0$ is three orders-of-magnitude higher than that in rhombohedral graphene devices with $N$ = 2, 3 and >10 at the same gating conditions (see Extended Figure 3). This is contradictory to the single-particle picture, which predicts a DOS that is >50 larger in pentalayer graphene than that in $N$ = 2 and 3. This experimental observation provides a direct and strong evidence of electron correlation effects in the zero-energy flat bands of pentalayer rhombohedral graphene.

Next, we explore the CI state in more detail. Figure 2a shows $R_{xx}$ at $n = 0$ and a range of $D$ in the temperature range of 5-100 Kelvin. At high $D$, $R_{xx}$ decreases when the temperature is increased, as expected for a BI state. At zero $D$, $R_{xx}$ follows the same trend, because the CI state is thermally excited. At high enough temperatures (see the inset and Supplementary Fig. S6), the state at zero $D$ becomes even more conducting than those at neighboring $D$s. This change indicates the melting of the CI state into a semimetal with large DOS as shown in Fig. 1b, and the recovery of the single particle picture.

The quantitative dependences on $T$, however, are different for BI and CI states as revealed in Fig. 2b. At $D = \pm 0.35$ V/nm, $R_{xx}$ follows an exponential dependence on ($1/T$) in the high-temperature range as indicated by the dashed line, implying a typical thermal activation behavior of band insulators. The values of gap $E_g$ extracted from this fitting is shown in the inset of Fig. 2b, where $E_g$ increases from almost zero at the gap re-opening $D$ to ~20 meV at higher $D$. In contrast, the state at $D = 0$ shows distinct behaviors at low and high temperatures. At low temperatures ($T < 40$K, blue region), $R_{xx}$ at $D = 0$ is several times higher than those at $D = \pm 0.35$ V/nm, and three orders-of-magnitude higher than those at $D = \pm 0.15$ V/nm. The $D = 0$ state clearly indicates an insulating behavior. At high temperatures ($T > 40$K, orange region), however, the $D = 0$ state has the same behavior as the semi-metallic state at $D = \pm 0.15$ V/nm, in contrast to the thermal activation behavior in the $D = \pm 0.35$ V/nm gapped state. The transition from insulator-like to metal-like behavior of the state at $D = 0$ clearly differentiates it from the states at $D = \pm 0.35$ V/nm. We conclude that the insulating behavior of the $D = 0$ state is due to a gap opened by the electron correlation effect, while at high temperatures this gap disappears so the device shows low resistance as predicted by the large DOS in the single-particle picture.

Theories have explored such Fermi surface instability and several candidates of a correlated insulating state at $D = n = 0$ have been suggested[37–44]. We examine the possibility of quantum valley Hall (QVH), quantum spin Hall (QSH), quantum anomalous Hall (QAH) and layer-anti-ferromagnetic (LAF) states[19] (see Supplementary Section III). The QVH state, which is layer polarized, can be ruled out as an explanation of our observation, as the device shows a symmetric gap closing and reopening behavior with respect to positive and negative $D$ as shown in Fig. 2a. The QSH and QAH states can also be ruled out, as the two-terminal conductance at $D = n = 0$ is close to zero, indicating the absence of edge states. As a result, the only candidate compatible with our observation is the LAF state. The definitive evidence of LAF, however, calls for in-depth spectroscopy experiments which can probe the whole band structure at $n = 0$ while $D$ is continuously tuned. This is beyond our current manuscript.

One trivial origin of the insulating state at $D = 0$ is a band insulator with some built-in interlayer potential difference. However, that state should evolve asymmetrically as $D$ is tuned to positive and negative values, in particular that low-resistance states should not appear on one side of $D$. In contrast, our data shown in Fig. 2a shows gap-closing behaviors at both positive and negative $D$s and mostly a symmetric dependence on $D$. We therefore rule out the built-in potential as the correct explanation of our observation.

**Correlation-driven Chern insulator states**

We examine the range of $D = 0.1$-$0.2$ V/nm in which our device shows a resistance on the order of several k$\Omega$. Figure 3a shows 2D color plots of $R_{xx}$ and $R_{xy}$ at $B = 1$ T. Large signals appear at around $n = 0$, due to the BI and CI states. In addition, three states emerge on the hole-doped side at $D = 0.11$, $0.16$ and $0.21$ V/nm, as indicated by the dashed lines. They start to emerge at 1 T and become fully developed at 2 T, as circled out in Fig. 1f. From the bottom to top on both plots, the first state is located at the same $D$ as where the CI state vanishes, the second state is located at where the system is metallic, while the third state is located at the same $D$ as where the BI state starts to develop---implying their close relation to the insulator-metal-insulator transitions at $n = 0$ when $D$ is continuously tuned. To further investigate these states, we first show in Fig. 3b the linecuts in Fig. 3a at $D = 0.16$ V/nm. The $R_{xy}$ is quantized at $\frac{h}{5e^2}$ and $R_{xx}$ shows a dip at $n = -1.2*10^{11}$ cm$^{-2}$. These observations indicate a Chern number $C = -5$. Figure 3c&d shows the evolution of $R_{xx}$ and $R_{xy}$ at $D$ = 0.16 V/nm as $B$ is tuned. The dashed lines trace this state and their slopes agree with $C$ = -5 according to Streda's formula[45]. Similarly, Fig. 3e shows the linecuts of $R_{xx}$ and $R_{xy}$ at $D$ = 0.11 V/nm and $B$ = 2 T, where $R_{xy}$ is quantized at $\frac{h}{3e^2}$ and $R_{xx}$ shows a dip at $n = -1.5*10^{11}$ cm$^{-2}$. Figure 3f&g show the evolutions of $R_{xx}$ and $R_{xy}$ at $D$ = 0.11 V/nm as $B$ is tuned. Both the quantized value of $R_{xy}$ and the slopes indicate a state with $C$ = -3. The state at $D$ = 0.21 V/nm also shows a Chern number $C$ = -3, as shown in Extended Data Fig. 6. For all the three states, the plateaus of quantized $R_{xy}$ are significantly broader than those of the states on the electron-doped side.

We understand the $C$ = -5 state as a correlation-driven Chern insulator state. The bands at zero energy in pentalayer rhombohedral graphene host a Berry phase of π per spin per valley. As the Berry curvature is concentrated at the band edge when a small gap is opened by the gate electric field, we can integrate it for the lower energy portion of the band and define a valley Chern number. The valley Chern number is 5/2 or –5/2 as shown in Fig. 3h, which evolves with the displacement field and allows for direct visualizing the physical picture. We start from the LAF state at zero $D$. The net Chern number of the system is zero due to the equal number of valence bands with positive and negative $C$. As $D$ is increased, the applied electrostatic potential will enlarge the gap in two of these isospin copies and shrink the gap in the other two copies, depending on their layer polarizations (Extended Fig. 9). When $D$ is large enough to close and re-open one of the gap-shrunk isospin-polarized bands, the corresponding Chern number will change sign[19,38]. As a result, the net Chern number of the system will be $C$ = -5. Such a QAH state at $B$ = 0 has been predicted by theory[19,38], but the experimental evidence with quantized $R_{xx}$ and $R_{xy}$ has not been observed. In our device, a finite magnetic field of ~0.9 T starts to reveal the $C$ = -5 Chern insulator state, which is aligned with the expectation that the QAH state might win the competition with the help of a small magnetic field[38]. The actual ground state at $B$ = 0 in our device could be a semi-metal, as we will discuss in the next section.

We also attribute Chern insulators as the origin of the $C$ = -3 states, which appear at values of $D$ that are roughly symmetric about the $D$ at which the $C$ = -5 state is observed. At the same time, these two values of $D$ define the boundaries between the CI state, the metallic state and the BI state. Both facts imply that the gap of one isospin band structure is closed at these $D$s (so is the global gap of the whole device). In these scenarios, counting the total Chern number is more complicated than for the $C$ = -5 state, as the subtlety in the band structure and the integration of Berry curvature over occupied states clearly play an important role[20,46,47]. Compared to the $C$ = -5 state which can be explained by the correlation-driven effects applied to isospin-polarized band structure at zero fillings, a band picture to explain the $C$ =

-3 states is not clear at this point. We have included more rationale against the integer quantum Hall states as explanations for the $C$ = -5 and $C$ = -3 states in Section VIII of Extended Data.

Similar Chern insulators were observed at the negative $D$ side (Extended Fig. 8). Considering the large twist angles between graphene and any hBN flakes, the moiré is negligible for both the top and bottom graphene/hBN interfaces. Therefore, the phase diagrams of the positive and negative $D$s are symmetric. Our observations invite further theoretical and experimental efforts for a complete understanding of the details.

**Competing semi-metallic phase at zero $B$ field**

The $C$ = -5 state vanishes at $B$ = 0.9 T, with an abrupt change of $R_{xx}$ and $R_{xy}$. This indicates the existence of a competing phase with the $C$ = -5 QAH state, instead of a disorder- or temperature-limited observation of the latter. Figure. 4a shows $R_{xx}$ in the phase transition range of $D$ at $n$ = $B$ = 0 and $T$ = 100 mK. Residing between the CI and BI states which both reach ~M$\Omega$, the competing phase shows a resistance of ~600 $\Omega$. Furthermore, the magneto-resistance at $n$ = 0 follows a quadratic dependence on $B$ as shown in the inset, which is typical for a compensated semi-metal. These observations indicate the competing phase to be a semi-metal, which is consistent with the single particle band structure of pentalayer graphene as shown in Fig. 1c. We further analyze in Fig. 4b the derivative of $R_{xx}$ versus $B$, which reveals a Landau fan of electron states converging to a negative electron density of 3.4*10$^{11}$ cm$^{-2}$. This fact agrees with the compensated semi-metal picture[48]. We note that the hot spot at $B$=0.9 T and $n$=-1.1*10$^{11}$ cm$^{-2}$ corresponds to the abrupt phase transition from the $C$=-5 CHI to the semi-metal, while the Landau levels on the electron-doped side disappear gradually.

The temperature-dependent $R_{xx}$ of this competing phase, however, is quite unusual. As shown in Fig. 4c, $R_{xx}$ increases as the temperature decreases in the range of $T$ = 7-40K. This is consistent with the semi-metal picture in which the density of thermally excited carriers is reduced as the temperature is decreased. At T < 7K, $R_{xx}$ decreases quickly as the temperature decreases. One might think this trend in the low-temperature range can be explained by reduced phonon-scattering as temperature is decreased. But our data at the large-$n$ metallic state rules out this explanation, as shown in Fig. 4c and Extended Fig. 10. This non-monotonic temperature dependence exists for a range of $D$, shown as the blue triangular region in the inset of Fig. 4c. The origin of this behavior is not clear, although the trend is intriguing: will $R_{xx}$ keep decreasing (even to zero eventually) as we further lower the temperature? Our observation invites further measurements at lower and well-calibrated electronic temperatures. The answer to this question could possibly shed light on the competing phase, as well as reveal other correlated and topological states such as superconductivity and QAH[19,49].

**Conclusions**

Our results demonstrated a collection of correlated and topological phases in a single material. These phenomena in a natural two-dimensional crystal demonstrate a pathway to explore exotic quantum phenomena without involving moiré superlattices. Recent theories suggest rhombohedral multilayer graphene as a topologically non-trivial chiral p+ip superconductor with chiral Majorana edge modes[50], and a valley-polarized Wigner crystal[51]. Our work establishes rhombohedral multilayer graphene as a highly tunable platform to study strongly correlated and topological states. The very different behavior of pentalayer graphene from thinner and much thicker graphene flakes also points to the layer thickness as an important knob to tune in further experiments on rhombohedral graphene.

**Acknowledgments**

We acknowledge helpful discussions with F. Zhang, T. Senthil, L. Levitov, L. Fu, Z. Dong, and A. Patri. L.J. acknowledges support from a Sloan Fellowship. Work by T.H. was supported by NSF grant no. DMR-2225925. The device fabrication of this work was supported by the STC Center for Integrated Quantum Materials, NSF grant no. DMR-1231319. Device fabrication was carried out at the Harvard Center for Nanoscale Systems and MIT.Nano. Part of the device fabrication was supported by USD(R&E) under contract no. FA8702-15-D-0001. K.W. and T.T. acknowledge support from the JSPS KAKENHI (Grant Numbers 20H00354, 21H05233 and 23H02052) and World Premier International Research Center Initiative (WPI), MEXT, Japan. H.P. acknowledges support by NSF grant no. PHY-1506284 and AFOSR grant no. FA9550-21-1-0216. A portion of this work was performed at the National High Magnetic Field Laboratory, which is supported by National Science Foundation Cooperative Agreement No. DMR-2128556* and the State of Florida.

**Author Contributions**

L.J. supervised the project. TH.H., Z.L., G.S., J.S. and J.W. performed the DC magneto-transport measurement. TH.H. and TY.H. fabricated the devices. K.W. and T.T. grew hBN single crystals. H.P. contributed to data analysis. All authors discussed the results and wrote the paper.

**Competing Interests** The authors declare no competing interests

**Figures:**

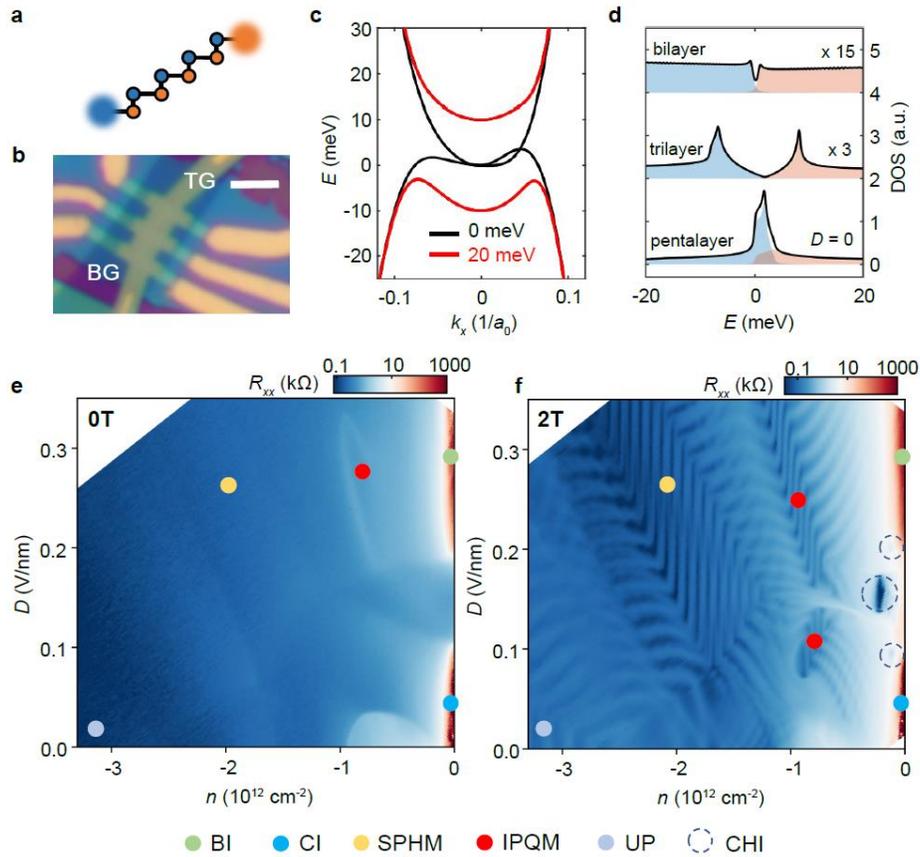

**Fig.1 | Correlated driven insulator, isospin symmetry broken states and Chern insulators in rhombohedral-stacked pentalayer graphene. a,** Schematic of rhombohedral-stacked pentalayer graphene. The wavefunctions of states in the lowest energy bands concentrate at the sublattices highlighted by the larger dots. **b,** Image of the device where the top gate (TG) and bottom gate (BG) are labeled. Scale bar: 3um. **c**, Tight-binding calculations of energy dispersion of rhombohedral-stacked pentalayer graphene under an interlayer potential $\Delta$ = 0 meV (black) and 20 meV (red) near the K point. With $\Delta$ = 0, metallic transport behaviors are expected at charge neutrality. **d,** Tight-binding calculation of density of states in Bernal bilayer, rhombohedral trilayer (time a factor of 15 and 3 for comparison), and pentalayer graphene at $D$ = 0. The blue and orange shaded areas depict the DOS of the valence and conduction band.  **e & f,** Color plots of four-probe resistance $R_{xx}$ as a function of carrier density $n$ and displacement field $D$ measured at $B$ = 0 and an out-of-plane $B$ = 2T at a temperature of 100 mK. Colored dots label different phases including band insulator (BI), correlated insulator (CI), spin-polarized half metal (SPHM), isospin-polarized quarter metal (IPQM), unpolarized metal (UP) and Chern insulator (CHI). The CI state signifies the strong electron correlation in pristine pentalayer graphene, which is beyond the single-particle picture in **c** & **d**.

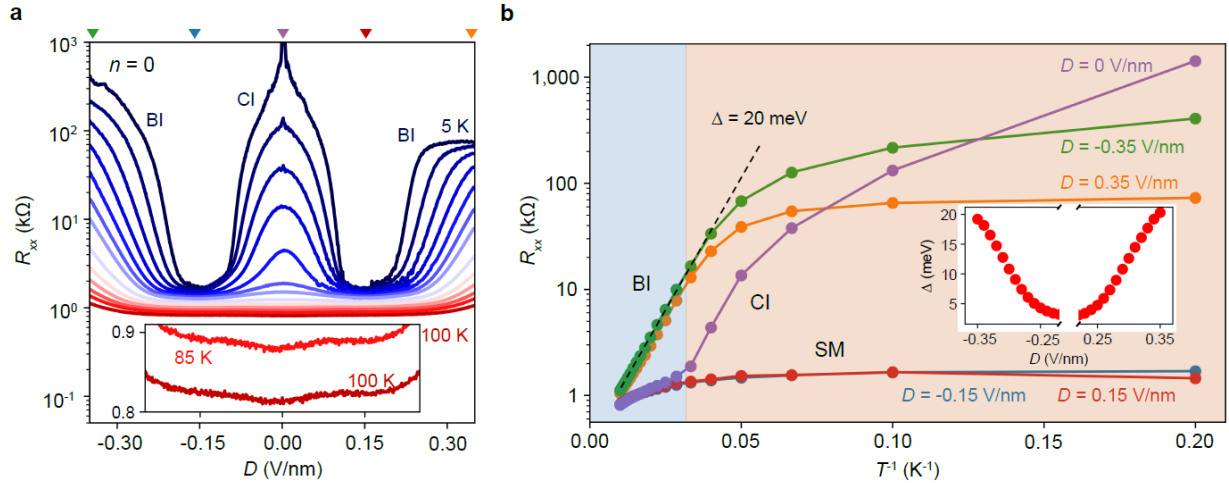

**Fig.2 | Temperature and magnetic field dependence of the correlated insulator state. a**, Temperature dependence of the four-probe resistance $R_{xx}$ measured at charge neutrality ($n = 0$) from 5K to 100K. Inset: Zoom-in of $R_{xx}$ versus $D$ at high temperatures 85K and 100K. At high temperatures the resistive state at $D = 0$ disappears and evolves to a dip in $R_{xx}$, consistent with the large single-particle density of states at D=0. **b**, $R_{xx}$ versus temperature from 5K to 100K at five displacement fields, corresponding to the colored triangles in **a** with the same color. They can be categorized into three groups, correlated insulator (CI), band insulator (BI) and semimetal (SM). At high temperatures (*T*>40K, orange region), the CI state resembles the SM state while at low temperatures (blue region), $R_{xx}$ of CI state increases dramatically, indicating the gap opening. Inset: bandgap of the band insulator extracted from the thermal activation fitting.

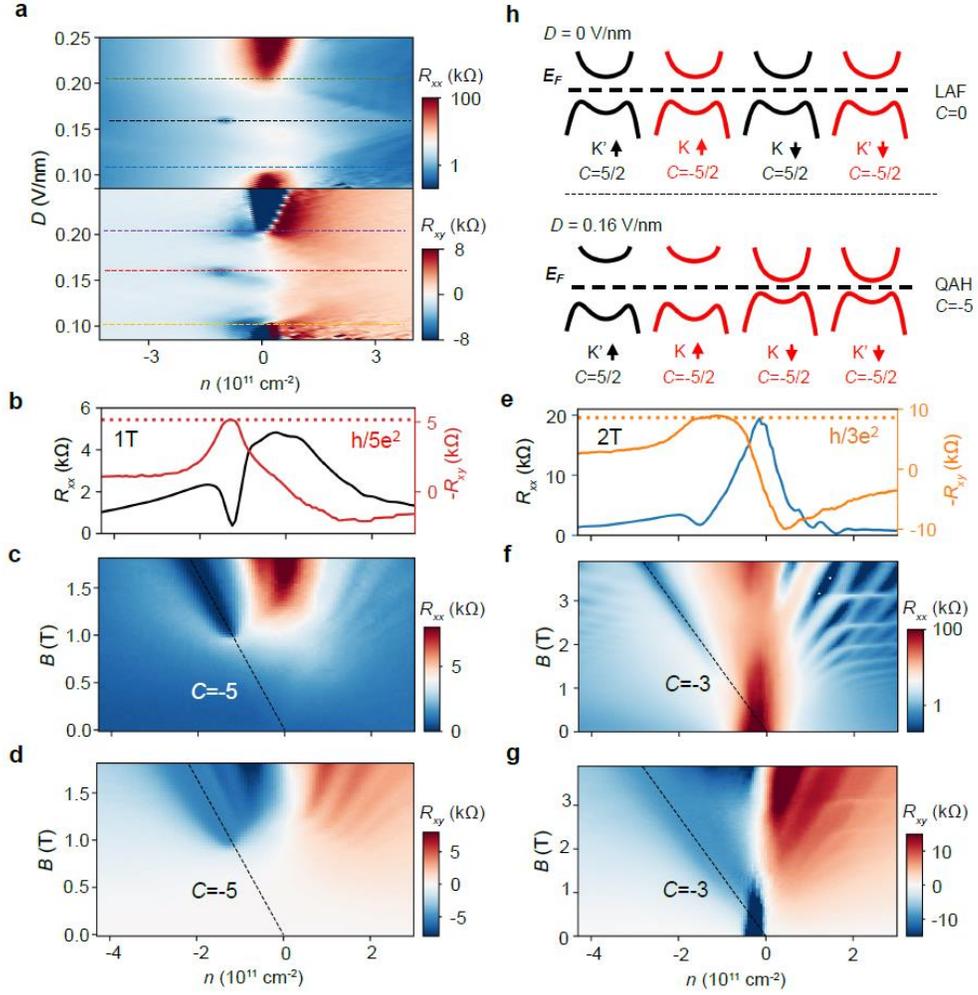

**Fig.3 | Correlation-driven Chern insulator states. a**, 2D color plots of $R_{xx}$ (upper panel) and $R_{xy}$ (lower panel) at $B$ = 1 T, revealing three Chern insulator states at the hole-doped side in the gap-closing range of $D$. The state with a Chern number $C$ = -5 happens at $D$ = 0.16 V/nm, while two states with $C$ = -3 happen at $D$ = 0.11 V/nm and $D$ = 0.21 V/nm, as indicated by the dashed lines. No signature of Chern insulator was observed on the electron-doped side. **b**, $R_{xx}$ and $R_{xy}$ corresponding to the dashed line at $D$ = 0.16 V/nm in **a**. $R_{xx}$ shows a significant dip at the charge density where the $R_{xy}$ is well quantized at $-h/5e^2$—indicating a Chern insulator with $C$ = -5. **c & d**, 2D color plots of $R_{xx}$ and $R_{xy}$ versus $n$ and $B$ at $D$ = 0.16 V/nm. The dashed lines indicate the $n$-$B$ relation of a $C$ = -5 Chern insulator as predicted by the Streda's formula. Compared to the faint Landau fan emerging at the electron-doped side, the $C$ = -5 state is much more robust and is the only state appearing on the hole side. Below $B$ = 0.95 T the $C$ = -5 state terminates sharply, indicating the existence of a competing phase. **e-g**, same plots as **b-d**, with $D$ = 0.11 V/nm and $C$ = -3. **h**, Schematic of the layer-antiferromagnetic state (LAF) at $D$ = 0 and the quantum anomalous Hall state (QAH) at an intermediate $D$, where the valleys and spins are labeled by (K, K') and arrows respectively. The color indicates the sign of the Chern number of the valence band of a specific isospin flavor. The net Chern number is 0 at $D$ = 0, while it becomes -5 as the bandgap corresponding to one isospin inverts upon applying a displacement field.

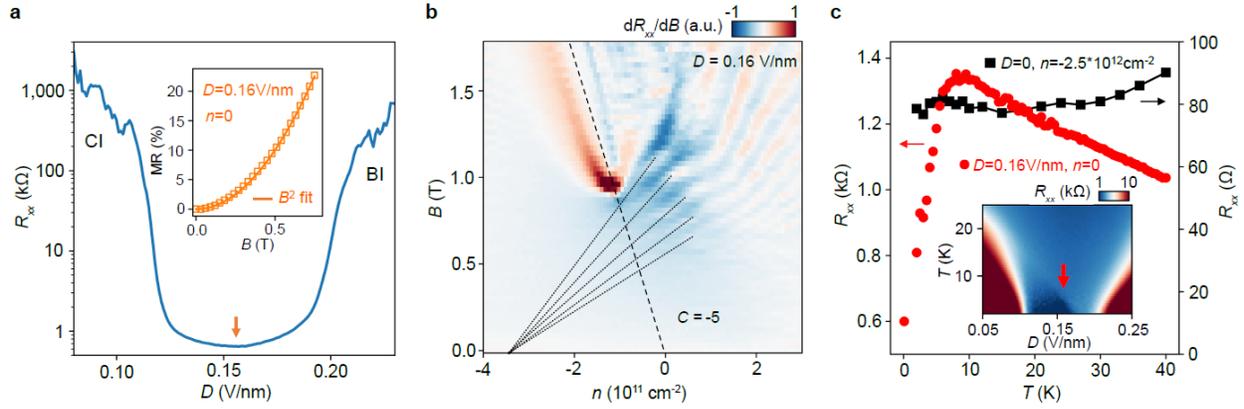

**Fig.4 | The competing phase at $B = 0$. a**, $R_{xx}$ versus $D$ at $n = 0$ and $T = 100$ mK. While $R_{xx}$ reaches ~MΩ at small $D$ (correlated insulator) and the large $D$ (band insulator), it goes down to well below kΩ in the intermediate range of $D$. Inset: magnetoresistance (MR) ratio $[R_{xx}(B) - R_{xx}(0)]/R_{xx}(0)$ at $D = 0.16$ V/nm and $n = 0$ as a function of $B$. The MR data (yellow squares) follows nicely a $B^2$ dependence, indicating a compensated semi-metal phase. **b**, Color plot of the $dR_{xx}/dB$ versus carrier density $n$ and out-of-plane magnetic field $B$ taken at $D=0.16$V/nm at a temperature of 100mK. The dashed line indicates the C=-5 state. The dotted lines trace the electron Landau levels and converge to a negative electron density of $3.4*10^{11}$ cm$^{-2}$, agreeing with the semimetal picture. **c**, Temperature-dependent $R_{xx}$ at $D = 0.16$ V/nm and $n = 0$ (red) and $D = 0$ and $n = -2.5*10^{12}$cm$^{-2}$ (black). The former curve increases as $T$ decreases at high temperatures but starts to drop quickly when $T$ is below 7 K, while the latter does not show strong temperature dependence at low temperatures. Inset: color map of $R_{xx}$ as a function of displacement field $D$ and temperature $T$ measured at $n = 0$. The dark blue region at low temperatures hosts the nonmonotonic temperature dependence and the red arrow indicates the $D = 0.16$ V/nm linecut.

# Methods and Extended Data

## I. Identification of rhombohedral pentalayer graphene and sample fabrication

The pentalayer graphene and hBN flakes were prepared by mechanical exfoliation onto $SiO_2$/Si substrates. As shown in Extended Data Figure 1, the rhombohedral domains of pentalayer graphene were identified using near-field infrared microscopy[52], confirmed with Raman spectroscopy and isolated by cutting with a Bruker AFM[53]. The van der Waals heterostructure was made following a dry transfer procedure. We picked up the top hBN and pentalayer graphene using polypropylene carbonate (PPC) film and landed it on a prepared bottom stack consisting of an hBN and graphite bottom gate. We intentionally misaligned the straight edges of the flakes to avoid the formation of the moiré superlattice. The device was then etched into a Hall bar structure using standard e-beam lithography (EBL) and reactive-ion etching (RIE). We further deposited a NiCr alloy top layer to form a dual-gate device.

## II. Transport measurement

The device was measured in a Bluefors LD250 dilution refrigerator with an electronic temperature around 100 mK. Stanford Research Systems SR830 lock-in amplifiers were used to measure the longitudinal and Hall resistance $R_{xx}$ and $R_{xy}$ with an AC voltage bias 200 uV at a frequency at 17.77 Hz.

## III. Tight-binding calculation

The single-particle band structure of the rhombohedral $N$-layer graphene is calculated from the 2N-band continuum model. The Hamiltonian and the parameters are taken from previous literature[18].

Extended Data Fig. 2 shows the calculated band structure and DOS of rhombohedral multilayer graphene (layer number $N$=2, 3, 4, 5, 7, 9) at zero interlayer potential ($\Delta$=0). Due to remote hoppings, the band structure deviates from $E \sim k^N$ at low energy. For bilayer and trilayer graphene, the DOS at zero energy is small due to the approximate $E \sim k^N$ dispersion. For rhombohedral graphene around 5 layers, there is a large DOS at zero energy, which leads to the Fermi surface instability and the correlated insulator state. For graphene thicker than 5 layers, the band-overlap and screening effect become larger, which weakens the interaction effect. This may explain the absence of the correlated insulator at $D = n = 0$ for thick rhombohedral graphene[22]. As a result, the rhombohedral pentalayer graphene is probably with the optimal layer number where the low energy bands are the flattest and the interaction is the strongest among all layer numbers.

## IV. Correlated insulator in pentalayer graphene

We show calculated and experimental results from bilayer and trilayer graphene to support the conclusion of a correlated insulator state in pentalayer graphene. In Extended Data Figure 3a, we present the DOS from the tight-binding calculation at $D = 0$ for layer number $N = 2$, 3 and 5 respectively. In Extended Data Figure 3b - 3d we present the resistance as a function of $n$ and $D$ for our devices with $N = 2$, 3 and 5. At $D = n = 0$, both $N = 2$ and 3 shows small DOS and ~ few kOhm resistance which is typical for metallic states in graphene. In contrast, the DOS at $D = n = 0$ for $N = 5$ is more than 50 times bigger than its counterparts in the cases of $N = 2$ and 3. In a single particle picture, such a large DOS should result in a much smaller resistance in $N = 5$ than its counterparts in $N = 2$ and 3. However, the experimentally measured value of

$R_{xx}$ > 10 MOhm is ~4 orders-of-magnitude bigger than those in $N$ = 2 and 3. This obvious contradiction to the single-particle picture clearly points to electron correlation effects as we concluded.

### V. Phase diagram for both electron and hole doping

Extended Data Fig. 4a and 4b show the color plots of four-probe resistance $R_{xx}$ as a function of carrier density $n$ and displacement field $D$ for the hole doping side and electron doping side measured at $B$ = 0 and a temperature of 100 mK. Colored dots label different phases including band insulator (BI), correlated insulator (CI), spin-polarized half metal (SPHM), isospin-polarized quarter metal (IPQM) and unpolarized metal (UP). The large $D$ side of the phase diagram looks similar to that of the previous study in bilayer and trilayer graphene. At the intermediate and small $D$ side, the semimetal and correlated insulator states emerge and dominate the neighboring regions.

The degeneracy of the isospin-polarized metal is inferred from the degeneracy of the Landau levels as described in the main text. However, the exact nature of the half metal requires further measurement. Extended Data Fig. 4c and 4e show the Hall resistance $R_{xy}$ as a function of the out-of-plane magnetic field at the red and yellow dot in 4a. The quarter-metal (4c) shows a clear anomalous Hall effect and magnetic hysteresis, indicating a net valley polarization. While the half-metal (4e) does not show anomalous Hall effect or magnetic hysteresis, indicating the absence of net valley polarization. Therefore, we conclude the half metal to be spin polarized and valley unpolarized.

### VI. Temperature dependence of the correlated insulator

We include a more complete set of temperature-dependent data of the correlated insulator. Extended Data Fig. 5a shows a complete data set of the temperature-dependent $R_{xx}$, in the range between 2K and 40K. The correlated insulator develops below 30K. The two white arrows indicate the semimetal phase with an anomalous temperature dependence in the low-temperature region, corresponding to Fig. 4c. Extended Data Fig. 5b shows the temperature dependence of $R_{xx}$ between 40K and 100K. The $D$ = 0 state shows as a bump in $R_{xx}$ at 40K and gets flatter as the temperature is increased. Eventually it becomes a dip at high temperatures. This evolution corresponds to the melting of the CI state into a metal with large DOS and the recovery of the single particle picture.

### VII. The $C$ = -3 Chern insulator state

The $C$ = -3 state at $D$ = 0.21 V/nm data is shown in Extended Data Fig. 6 and shows similar behavior to the $C$ = -3 state at $D$ = 0.11 V/nm shown in Fig. 3. Similar to Fig. 4b in the main text, we show the derivative plot of the $C$ = -3 state at $D$ = 0.11 V/nm in Extended Data Fig. 7. Based on the $dR_{xx}/dB$ in (7a) and $dR_{xy}/dn$ in (7b), we found the $C$ = -3 state is the only state at the hole side under low magnetic fields and traces all the way to zero magnetic field. The integer quantum Hall states on the electron side disappear at a higher magnetic field. The $C$ = -5 state has a competing semimetal phase at $B$ = 0, which dominates the state at right below 0.9 T. In contrast, there is no such a competing phase for the $C$ = -3 state, so the $C$ = -3 state traces all the way down to $B$ = 0.

### VIII. Rationale to rule out the integer quantum Hall state as an explanation for the $C$ = -5 state

Now we explain why the $C$ = -5 state cannot be an integer quantum Hall state. For rhombohedral-stacked graphene with layer number $N$, when considering only the nearest neighbor hopping, the integer quantum

Hall states near zero energy should first appear at $C = \pm 2N$. This is due to the $N$ almost-degenerate Landau levels at zero energy and the 4-fold degeneracy of spin and valley[17]. In our device, however, the $C = \pm 10$ states are completely missing. It is likely that the pseudospin degeneracy has already been lifted at $D = 0.16$ V/nm, and the observed $C = -5$ state corresponds to one of the four pseudospin copies of the band structure. But if we apply a simple integer quantum Hall picture to one pseudospin polarized band structure, one would expect a $C = 5$ state on the electron-doped side to appear at an even smaller $B$ field, due to the more dispersive conduction band and a larger quantum Hall gap correspondingly (see Fig. 1c). In addition, integer quantum Hall states corresponding to C = -6, -7, -8 and so on should appear at even lower $B$, as those gaps are bigger than that of the $C = -5$ state[17,20]. The experimental data, however, is contradictory to these expectations: the $C = 5, -6, -7, -8$ and other quantum Hall states are completely missing in Fig. 3a when the $C = -5$ state has fully developed with a quantized $R_{xy}$ value. This is distinct from in bilayer graphene, where the $C = \pm 2$ and $C = \pm 4$ states appear at a similar $B$ value[27,54].

Even if we include more hopping terms in the tight-binding model, the integer quantum Hall picture still cannot explain our data. In this case, the trigonal warping effect distorts the $E \propto \pm |k|^N$ energy dispersion and creates a 3-fold rotational symmetric valence band (see Supplementary Fig. S1)[18,20]. At close to zero energy, the integer quantum Hall states correspond to $C = -3m$, where the 3 is due to the three hole-pockets and the integer $m$ is the number of the isospin flavors at the Fermi level. Previous experiments on bilayer and trilayer graphene all show data consistent with this picture[23,29,36,55]. The $C = -5$ observed in our device, however, clearly deviates from this integer quantum Hall picture. At hole densities larger than $2.8*10^{12}$ cm$^{-2}$, the Fermi surface recovers to a single circle and the quantum Hall sequence no longer follows $C = -3m$ (see Supplementary Fig. S1). However, this density threshold is one order of magnitude bigger than $1*10^{11}$ cm$^{-2}$, the density at which we start to observe the $C = -5$ state[22,28,35].

We also emphasize that the $C = -3$ states are also likely Chern insulators for the following reasons. 1. There are no other states with finite Chern numbers on the hole-doped side. 2. The plateau of $R_{xy}$ for the $C = -3$ state is much wider than that of the states on the electron-doped side. 3. These states persist to a lower magnetic field than any other integer quantum Hall states at the same displacement field (see Fig. 3e and extended figures).

**Methods-Only References**

**Extended Data Figures**

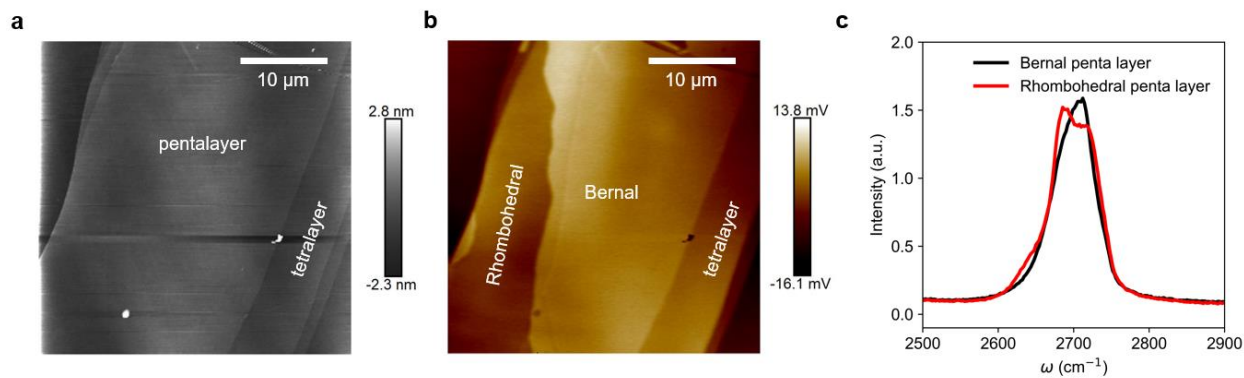

**Extended Data Fig. 1 | Identification of rhombohedral pentalayer graphene. a**, AFM topography map of a pentalayer graphene sample on a SiO$_2$/Si substrate. (The small region on the right corresponds to a graphene tetralayer.) **b**, Near-field infrared nanoscopy image of the same pentalayer graphene sample as in **a**, showing different contrast on the pentalayer region. The bright region corresponds to Bernal stacking and the darker region corresponds to rhombohedral stacking. **c**, Raman spectra taken at rhombohedral and Bernal stacking domains in the pentalayer graphene sample as in **b**. ω is the Raman shift.

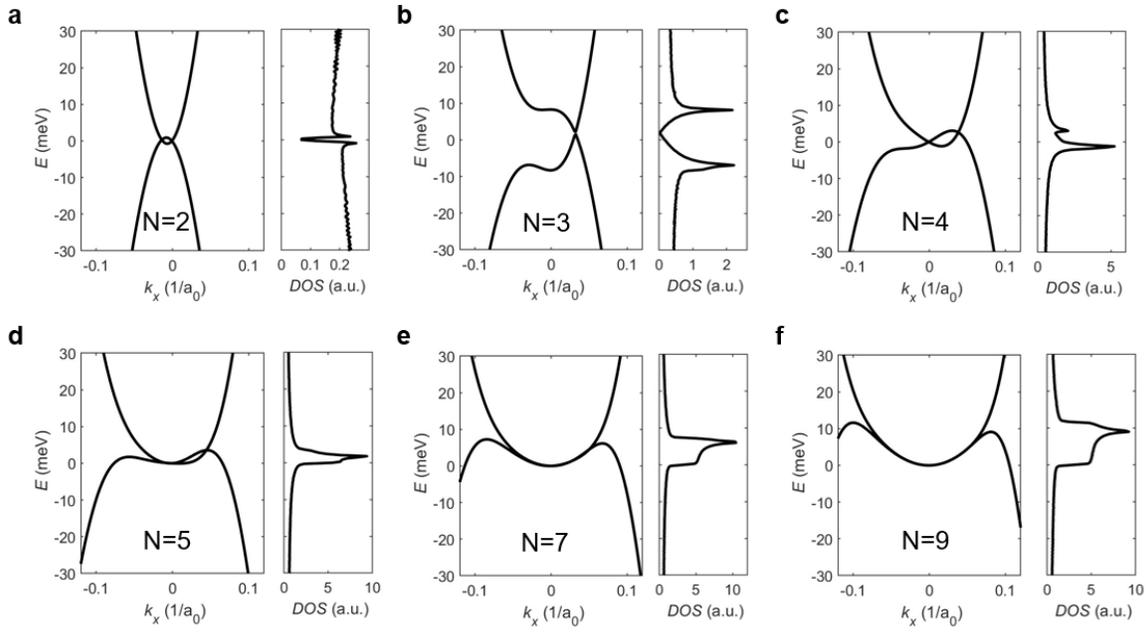

**Extended Data Fig. 2 | Single particle band structure and density of states of rhombohedral multilayer graphene. a-f,** Tight-binding calculation of single-particle band structure and density of states (DOS) for rhombohedral multilayer graphene (layer number $N$ = 2, 3, 4, 5, 7, 9). Due to the remote hopping terms, the band structure deviates from $E \sim k^N$ at low energy. The rhombohedral pentalayer graphene has the flattest band among all layer numbers.

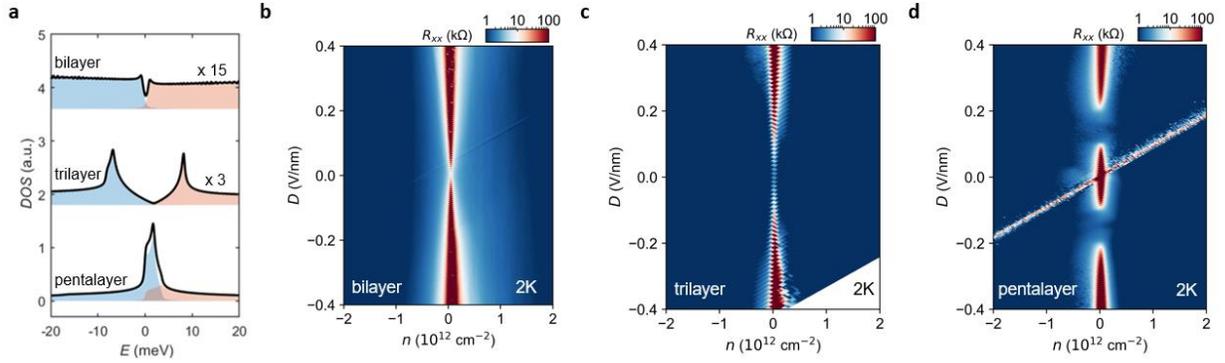

**Extended Data Fig. 3 | Correlated insulator at *n* = *D* = 0 in pentalayer rhombohedral graphene. a,** The calculated single-particle density of states (DOS) in Bernal bilayer, rhombohedral trilayer (time a factor of 15 and 3 for comparison), and pentalayer graphene at *D* = 0. The blue and orange shaded areas depict the DOS of the valence and conduction band. **b-d**, The *n-D* $R_{xx}$ map of the bilayer, trilayer and pentalayer graphene measured at 2K. Pentalayer graphene has a much larger DOS and band overlap at *n* = 0 compared to the bilayer and trilayer case and expects to be more conducting at *D* = *n* = 0. However, both bilayer and trilayer graphene are conducting at *D* = *n* = 0, while an insulating state appears at *D* = *n* = 0 in pentalayer graphene, indicating the non-single-particle origin of the insulating state. The off-diagonal line in the *n-D* map of pentalayer graphene is due to the big contact resistance when the bottom gate is zero.

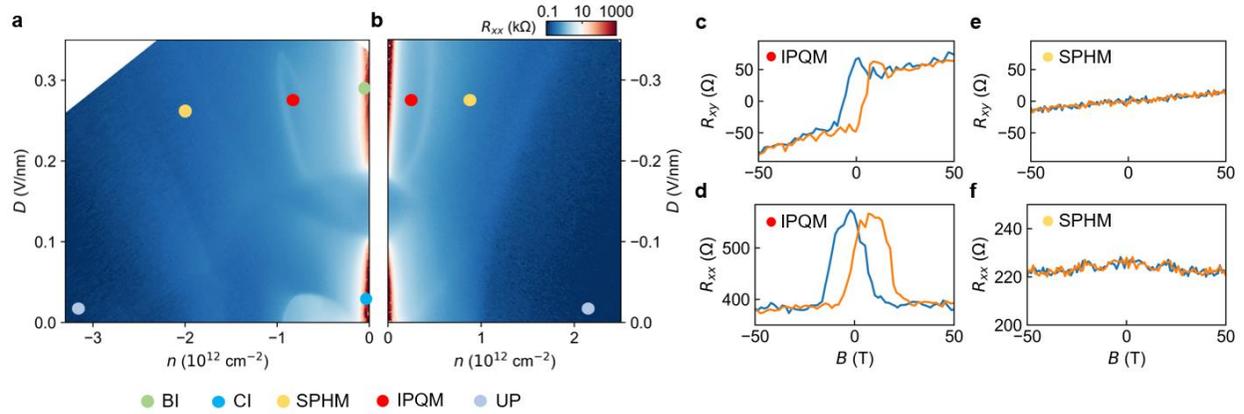

**Extended Data Fig. 4 | Phase diagram for both electron and hole doping. a, b,** Color plots of four-probe resistance $R_{xx}$ as a function of carrier density $n$ and displacement field $D$ for the hole doping side and electron doping side measured at $B = 0$ and a temperature of 100 mK. Colored dots label different phases including band insulator (BI), correlated insulator (CI), spin-polarized half metal (SPHM), isospin-polarized quarter metal (IPQM) and unpolarized metal (UP). **c, d,** Hall resistance $R_{xy}$ and longitudinal resistance $R_{xx}$ as a function of the out-of-plane magnetic field at the red dot in **a**. **e, f,** $R_{xy}$ and $R_{xx}$ at the yellow dot in **a**. The quarter metal **c & d** shows a clear anomalous Hall effect and magnetic hysteresis, indicating a net valley polarization. While the half metal **e & f** does not show anomalous Hall effect or magnetic hysteresis, indicating the absence of net valley polarization. Therefore, we conclude the half metal to be spin polarized but valley unpolarized.

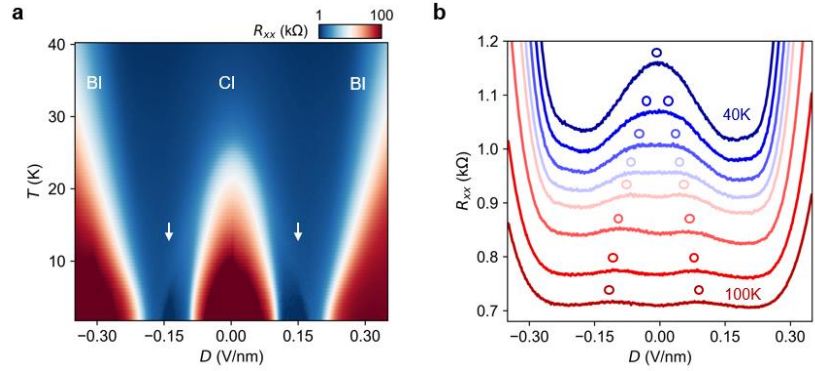

**Extended Data Fig. 5 | Additional temperature dependence data of the correlated insulator. a**, Temperature dependence of the four-probe resistance $R_{xx}$ measured at charge neutrality ($n = 0$). BI and CI stand for band insulator and correlated insulator. The correlated insulator develops below ~ 30K. The two white arrows indicate the semimetal phase with an anomalous temperature dependence at the low-temperature region, discussed in Fig 4c. **b**, $R_{xx}$ versus $D$ at even higher temperatures. As the temperature is increased, the resistive state at $D = 0$ disappears and evolves to a dip in $R_{xx}$. Circles trace the center of the bumps in $R_{xx}$.

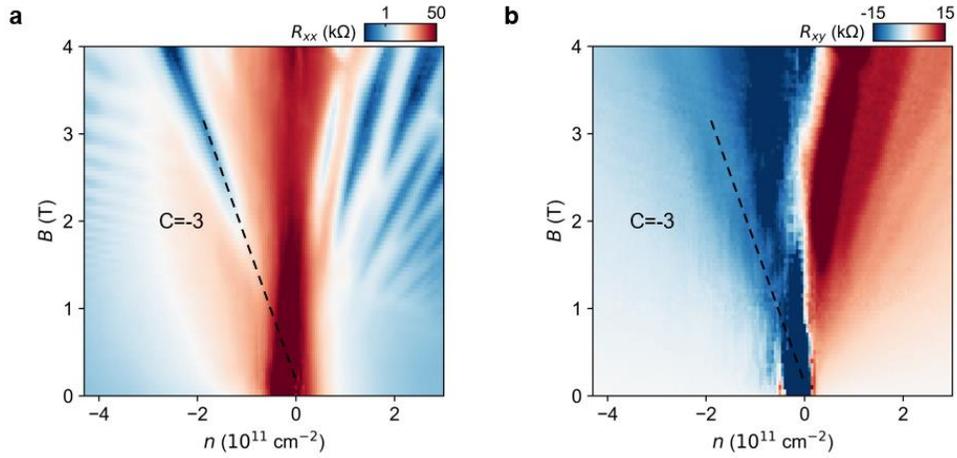

**Extended Data Fig. 6 | The *C* = -3 state at *D* = 0.21V/nm**. **a, b,** 2D color plot of $R_{xx}$ and $R_{xy}$ versus carrier density *n* and out-of-plane magnetic field *B* taken at *D* = 0.21V/nm at a temperature of 100mK. The dashed line indicates the *C* = -3 state. The *C* = -3 state is the only visible state on the hole doping side at low magnetic fields, in contrast to the electron side where all Landau levels appear at a similar magnetic field.

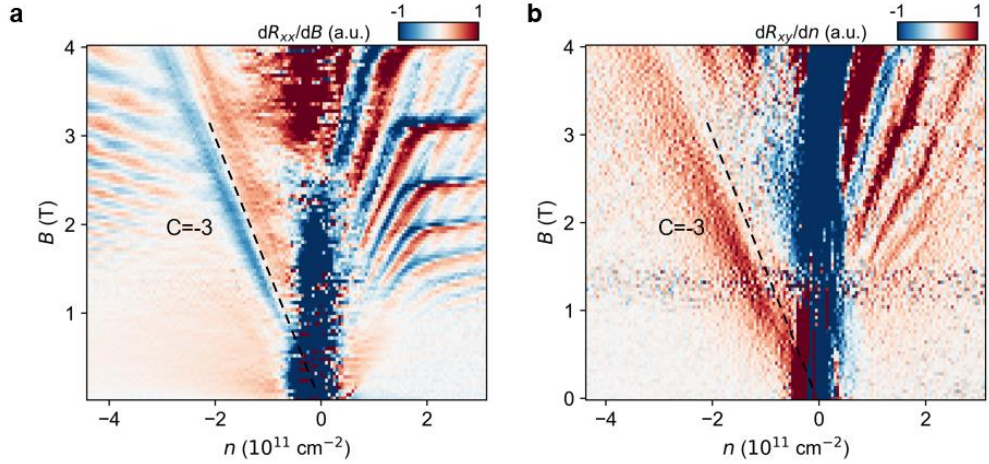

**Extended Data Fig. 7 | Tracing the *C* = -3 state towards zero magnetic field. a,** 2D color plot of d$R_{xx}$/d$B$ versus carrier density *n* and out-of-plane magnetic field *B* taken at *D* = 0.11V/nm at a temperature of 100mK. The dashed line indicates the *C* = -3 state. **b,** 2D color plot of d$R_{xy}$/d*n* versus carrier density *n* and out-of-plane magnetic field *B* taken at *D*=0.11V/nm at a temperature of 100mK. The dashed line indicates the *C* = -3 state. The *C* = -3 state is the only visible state on the hole doping side at low magnetic field and traces all the way to 0T. On the electron side, a complete set of Landau levels is observed and disappears at around 1.5T.

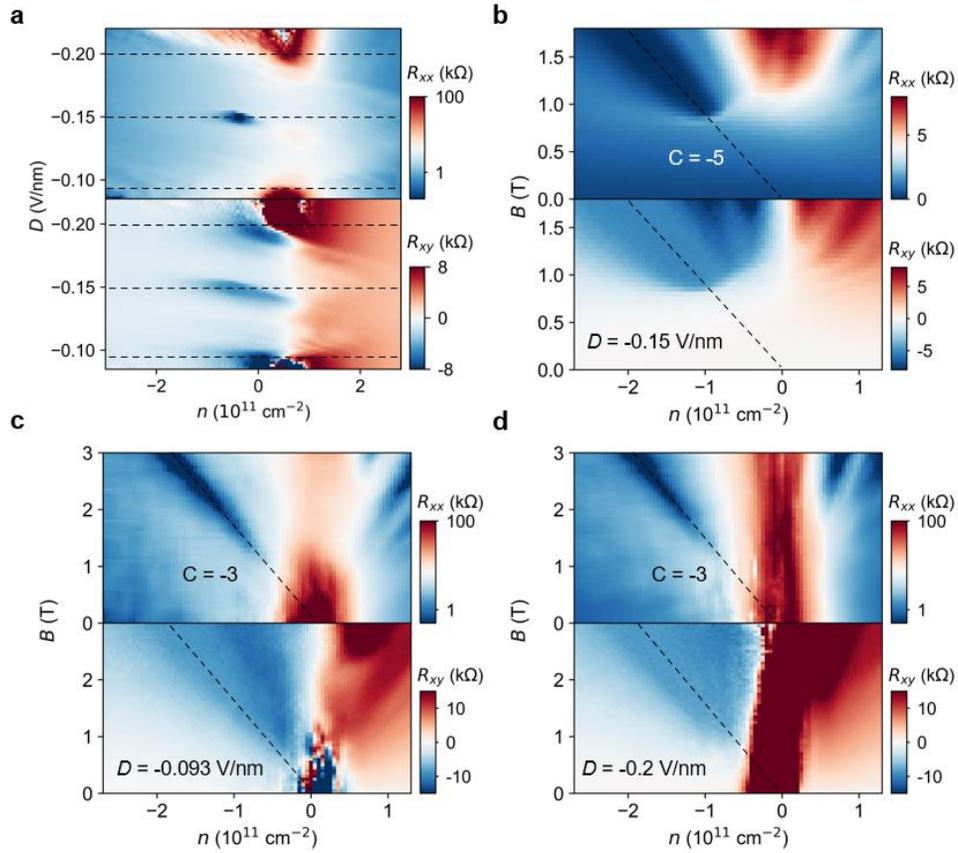

**Extended Data Fig. 8 | Chern insulators at the negative side of *D*. a**, 2D color plots of $R_{xx}$ (upper panel) and $R_{xy}$ (lower panel) at $B = 1$ T, revealing three Chern insulator states at the hole-doped side in the gap-closing range of *D*. The state with a Chern number $C = -5$ happens at $D = -0.15$ V/nm, while two states with $C = -3$ happen at $D = -0.093$ V/nm and $D = -0.2$ V/nm, as indicated by the dashed lines. **b, c, d**, 2D color plots of $R_{xx}$ (upper panel) and $R_{xy}$ (lower panel) versus *n* and *B* at $D = -0.15$ V/nm, $-0.093$ V/nm and $-0.2$ V/nm respectively. The dashed lines indicate the *n-B* relation of the Chern insulators as predicted by the Streda formula. These results are consistent with those of the positive *D* side.

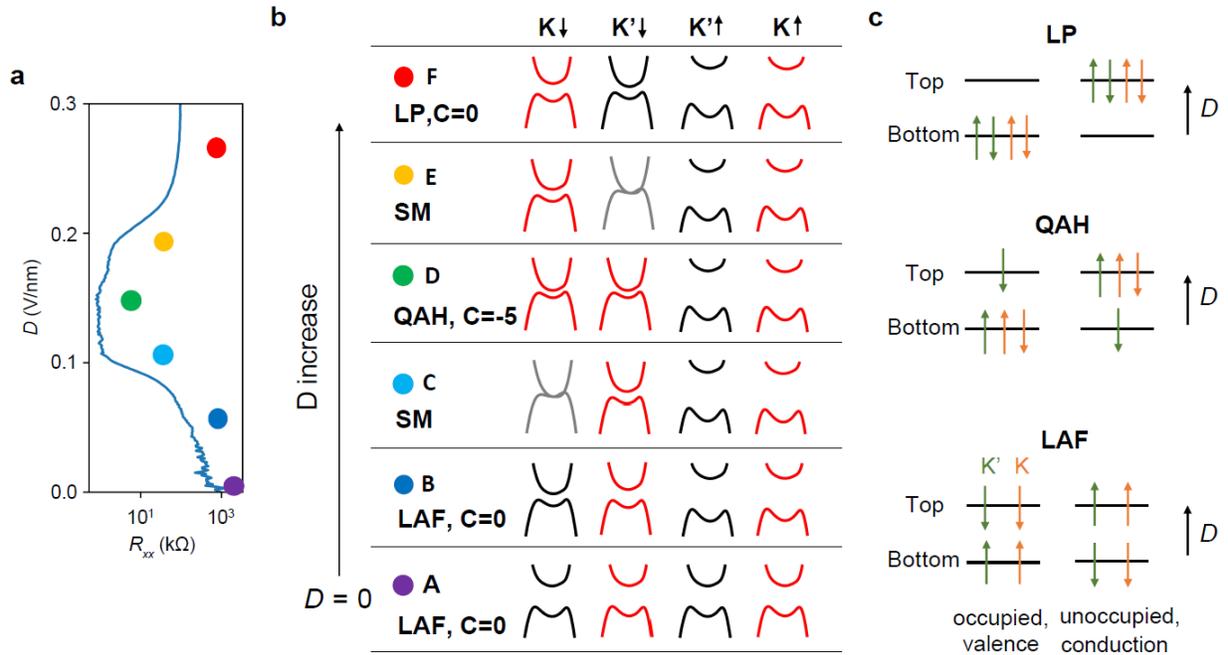

**Extended Data Fig. 9 | Evolution of the band structure at charge-neutrality with D. a,** $R_{xx}$ measured at $n$ = 0 at 2K. **b,** The band structure schematic of each isospin flavor at charge-neutrality under different $D$ values. The colored dots correspond to the states in **a**, including LAF (layer antiferromagnet), SM (semimetal), QAH (quantum anomalous Hall) and LP (layer polarized state). The color of each band represents the valley Chern number (black: 5/2, red: -5/2), while the band with a grey color indicates the gap-closing case. **c,** The layer polarization of the LP, QAH and LAF state. The layer polarization of both conduction and valence band for each isospin flavor is shown, where green and orange color corresponds to the K' and K valley. At $D$ = 0, the system starts with the LAF state where charges are evenly distributed in the top and bottom layer in a spin-polarized manner (point A). As $D$ increases, the gap of spin up (down) flavor expands (shrinks) due to its layer configuration (point B). It is important to note that the gap sizes of the two valleys with the same spin may differ. Consequently, one of the two gaps closes and reopens first (point C), leading to a situation where the gap of the $K$ valley inverts while the gap of the $K'$ valley remains the same, resulting in the so-called QAH state (point D). Moreover, the layer polarization becomes partially polarized at this stage. If $D$ continues to increase, the gap of the other valley eventually closes and reopens (point E), leading to the fully layer-polarized state (point F).

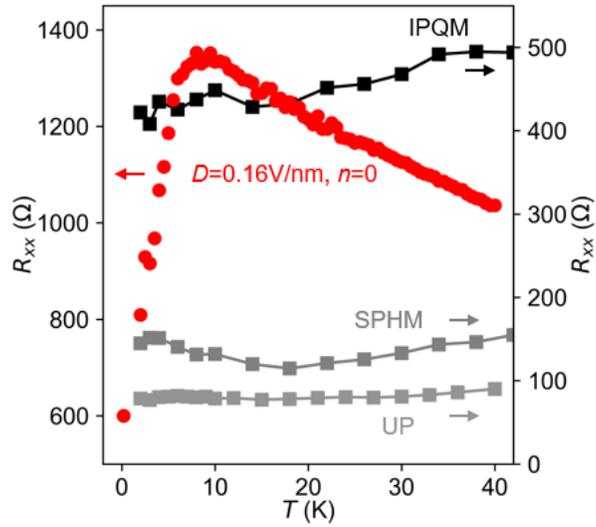

**Extended Data Fig. 10 | Temperature dependence of the SPHM, IPQM and UP states.** Temperature-dependent $R_{xx}$ at $D$ = 0.16 V/nm and $n$ = 0 (red), $D$ = 0 and $n$ = -2.5*$10^{12}$cm$^{-2}$ (light grey, UP), $D$ = 0.26 V/nm and $n$ = -1.9*$10^{12}$cm$^{-2}$ (dark grey, SPHM), and $D$ = 0.26 V/nm and $n$ = -0.8*$10^{12}$cm$^{-2}$ (black, IPQM).